\documentclass[conference]{IEEEtran}
\usepackage[T1]{fontenc}
\usepackage[latin9]{inputenc}
\usepackage{color}
\usepackage{float}
\usepackage{amsthm}
\usepackage[nosumlimits]{amsmath}
\usepackage{amssymb}
\usepackage{multirow}
\usepackage{overpic}
\usepackage{hyperref}
\usepackage{breakurl}

\makeatletter

\floatstyle{ruled}
\newfloat{algorithm}{tbp}{loa}
\providecommand{\algorithmname}{Algorithm}
\floatname{algorithm}{\protect\algorithmname}

\theoremstyle{plain}

\theoremstyle{definition}

\theoremstyle{plain}

\theoremstyle{remark}
\newtheorem{rem}{\protect\remarkname}
\theoremstyle{remark}

\allowdisplaybreaks

\usepackage{mathtools}

\usepackage{amsfonts}
\usepackage{cite}
\usepackage{array}
\usepackage{algorithm}
\usepackage{algorithmic}
\usepackage{graphicx}
\usepackage{subfigure}

\DeclareMathOperator*{\maxi}{max}

\DeclareMathOperator*{\st}{s.t.}

\newcommand{\herm}{^{H}}
\newcommand{\trans}{^{T}}
\renewcommand{\Im}{\mathrm{Im}}
\renewcommand{\Re}{\mathrm{Re}}

\newcommand*{\rom}[1]{\expandafter\@slowromancap\romannumeral #1@}

\makeatother

\providecommand{\definitionname}{Definition}
\providecommand{\factname}{Fact}
\providecommand{\remarkname}{Remark}
\providecommand{\theoremname}{Theorem}
\providecommand{\lemmaname}{Lemma}

\mathchardef\mhyphen="2D

\IEEEoverridecommandlockouts

\begin{document}
\title{Energy-Efficient Coordinated Multi-Cell Multigroup Multicast Beamforming with Antenna Selection}

\author{\IEEEauthorblockN{Oskari Tervo\textsuperscript{$\ast$}, Le-Nam Tran\textsuperscript{$\dag$}, Harri Pennanen\textsuperscript{$\ast$}, Symeon Chatzinotas\textsuperscript{$\ddagger$}, Markku Juntti\textsuperscript{$\ast$}, and Bj\"orn Ottersten\textsuperscript{$\ddagger$}}
\IEEEauthorblockA{\textsuperscript{$\ast$}Centre for Wireless Communications, University of Oulu, Oulu, Finland\\
\textsuperscript{$\dag$}Department of Electronic Engineering, Maynooth University, Maynooth, Co Kildare, Ireland\\
\textsuperscript{$\ddagger$}the Interdisciplinary Centre for Security, Reliability and Trust, University of Luxembourg, Luxembourg \\
Email: \{firstname.lastname\}@oulu.fi, lenam.tran@nuim.ie, \{symeon.chatzinotas, bjorn.ottersten\}@uni.lu}
}


\maketitle

\begin{abstract}
This paper studies energy-efficient coordinated beamforming in multi-cell multi-user multigroup multicast multiple-input single-output systems. We aim at maximizing the network energy efficiency by taking into account the fact that some of the radio frequency chains can be switched off in order to save power. We consider the antenna specific maximum power constraints to avoid non-linear distortion in power amplifiers and user-specific quality of service (QoS) constraints to guarantee a certain QoS levels. We first introduce binary antenna selection variables and use the perspective formulation to model the relation between them and the beamformers. Subsequently, we propose a new formulation which reduces the feasible set of the continuous relaxation, resulting in better performance compared to the original perspective formulation based problem. However, the resulting optimization problem is a mixed-Boolean non-convex fractional program, which is difficult to solve. We follow the standard continuous relaxation of the binary antenna selection variables, and then reformulate the problem such that it is amendable to successive convex approximation. Thereby, solving the continuous relaxation mostly results in near-binary solution. To recover the binary variables from the continuous relaxation, we switch off all the antennas for which the continuous values are smaller than a small threshold. Numerical results illustrate the superior convergence result and significant achievable gains in terms of energy efficiency with the proposed algorithm.

\end{abstract}


\setlength{\belowdisplayskip}{4.6pt}
\setlength{\abovedisplayskip}{4.6pt}
\setlength{\textfloatsep}{3pt }

\section{\label{sec:intro}Introduction}

The energy efficiency (EE) of the multi-antenna technologies is a performance indicator growing its importance for wireless communications and future 5G network design \cite{Isheden:Framework:EE:2012,Ng:EE:OFDMA:2012,Venturino-15,Jiang-11,Tervo-15,Nguyen-15}. The total power consumption consists of various elements, such as base stations (BSs) direct current (DC) power, baseband processing, radio frequency (RF) processing, coding, decoding, and backhaul operations. In fact, the data transmit power takes only a small share of the total power consumption when moving towards 5G networks \cite{Osseiran-14}. Each active antenna element requires an RF chain to preprocess the signal which has a significant impact on the power consumption. An efficient method to improve the energy efficiency is to use only some of the antennas for transmission by switching off a portion of the RF chains, giving rise to the problem of joint beamforming and antenna selection \cite{Jiang-Cimini-12,Tervo-15}.

Coordinated beamforming \cite{Irmer-11} has been one of the most efficient methods to deal with the inter-cell interference in the research on the multi-cell networks. The technique, adopted in the current LTE-A standards, uses the coordination of channel state information or inter-cell interference conditions to improve the system performance. Energy-efficient coordinated beamforming has been extensively studied recently, e.g., in \cite{Nguyen-15,Li-14CoMP,Tervo-16Arxiv,TervoWsumEE-15}.


The research presented before focuses mostly on unicast beamforming where each user is assigned with an independent data stream. However, the increasing demand for high data rate applications such as video broadcasting services creates new challenges. This gives rise to the concept of multicast beamforming, where multiple users desire to receive the same information. Multicasting is a particularly powerful technique in the context of cache-enabled cloud radio access networks proposed for 5G systems, where it can be used to transmit the same popular contents to multiple users to improve both spectral and energy efficiency \cite{Tao-15}. The physical layer multicasting has been also included in the LTE standards and it has applications in satellite communications \cite{Vazquez-16,Pen-16Sat} as well. Multicast beamforming problems have been studied in single-cell systems for different optimization targets, e.g., transmit power minimization \cite{Sidiropoulos-06,Karipidis-08}, max-min fairness \cite{Karipidis-08,Christopoulos-14}, and sum rate maximization \cite{FrameBased-15}. Joint beamforming and antenna selection for transmit power minimization was studied in \cite{Mehanna-13}.
Coordinated multicast beamforming for transmit power minimization and max-min fairness has been studied in \cite{Xiang-13}. He {\it et al.} \cite{He-15} proposed a centralized algorithm for the energy efficiency maximization problem in multi-cell system with single-group per cell. However, \cite{He-15} only considered the beamforming problem without taking into account the fact that significant energy efficiency savings can be achieved by switching off some of the RF chains, i.e., antenna selection.


In this paper, we study energy-efficient coordinated beamforming in multi-cell multigroup multi-user multicast multiple-input single-output (MISO) systems with joint beamforming and antenna selection. We first introduce a novel way to reduce the feasible set of the continuous relaxation, resulting in better performance compared to the original formulation. Then, by proper equivalent transformations, the continuous relaxation of binary variables, and the Charnes-Cooper transformation, we propose a successive convex approximation (SCA) based algorithm to solve the problem. Thanks to the proposed formulation, the continuous relaxation yields some continuous values being close to binary ones. To this end, we switch off antennas which have continuous values close to zero.

Contrary to \cite{He-15}, where an energy-efficient coordinated beamforming method for fixed antenna sets was derived with single group per cell, we focus on energy-efficient joint beamforming and antenna selection problem with multiple groups per cell which has not been considered in the previous multicasting research. In addition, we set antenna specific maximum power constraints whereas \cite{He-15} only considered the BS-specific sum power constraints which can result in large dynamic range of the PA output powers, causing large phase deviations due to phase non-linearities \cite{DVBTR15}.

The rest of the paper is organized as follows. Section \ref{sec:ProblemFormulation} presents the system model, power consumption model and the considered optimization problem. The proposed algorithms are provided in Section \ref{sec:CentralizedMethods} while numerical evaluation and conclusions are presented in Section \ref{sec:NumericalResults} and \ref{sec:Conclusions}, respectively.

The following notations are used in this paper. $|x|$ denotes the cardinality of $x$ if $x$ is a set, and absolute value of $x$, otherwise. $||\mathbf{x}||_2$ is the Euclidean norm of $\mathbf{x}$ and boldcase letters are vectors. $\mathbf{x}\trans, \mathbf{x}\herm, \Re(\mathbf{x})$ and $\Im(\mathbf{x})$ mean transpose, Hermitian transpose, real part and imaginary part of $\mathbf{x}$, respectively. \vspace{-1mm}

\section{\label{PF} System Model and Problem Formulation}
\label{sec:ProblemFormulation}
\subsection{System Model} \vspace{-2mm}
We consider a multi-cell multigroup multicasting system with $B$ BSs, where each BS $b \in \mathcal{B}=\{1,\ldots,B\}$, transmits independent information to $G_b$ groups of users in its cell. The set of groups served by BS $b$ is denoted by $\mathcal{G}_b=\{j,j+1,\ldots,j+G_b-1\}$. The total number of users and groups in the network is denoted by $K=|\mathcal{K}| (\mathcal{K}=\{1,\ldots,K\})$ and $G=|\mathcal{G}| (\mathcal{G}=\{1,\ldots,G\})$, respectively. BS $b$ is equipped with $N_b$ transmit antennas, whereas each user has only one receive antenna.
The serving BS of user group $g$ is denoted as $b_g$.
The set of users in group $g$ is denoted by $\mathcal{K}_g \subset \mathcal{K}$.
Since each user belongs to only one group, the sets of users belonging to different groups are disjoint, i.e., $\mathcal{K}_i \cap \mathcal{K}_j = \emptyset$, $\forall i,j \in \mathcal{G}, i \neq j$.
The received signal at user $k$ in group $g$ is given by
\begin{eqnarray} \label{eq:RxSignal}
{y}_{k} &=& \overbrace{\mathbf{h}_{b_g,k} \mathbf{w}_{g} {s}_{g}}^\text{desired signal} + \overbrace{\sum\limits_{i \in \mathcal{G}_{b} \setminus \{g\}} \mathbf{h}_{b_g,k} \mathbf{w}_{i} {s}_{i}}^\text{intra-cell interference} \nonumber \\
&& + \underbrace{\sum\limits_{j \in \mathcal{B} \setminus \{b_g\}} \sum\limits_{u \in \mathcal{G}_j} \mathbf{h}_{j,k} \mathbf{w}_{u} {s}_{u}}_\text{inter-cell interference} + {n}_{k}
\end{eqnarray}
where $\mathbf{h}_{b,k} \in \mathcal{C}^{1 \times N_b}$ is the channel (row) vector from BS $b$ to user $k$, $\mathbf{w}_{g} \in \mathcal{C}^{N_b \times 1}$ is the transmit beamforming vector of group $g$, ${s}_{g} \in \mathcal{C}$ is the corresponding normalized data symbol and ${n}_{k} \ \sim \mathcal{C} \mathcal{N} (0, \sigma_{k}^{2})$ is the complex white Gaussian noise sample with zero mean and variance $\sigma_{k}^{2}$. The SINR of user $k$ can be written as
\begin{equation}
\Gamma_k(\mathbf{w}) = \frac{|\mathbf{h}_{b_g,k} \mathbf{w}_{g}|^{2}}{{\sigma_{k}^{2} + \sum\limits_{u \in \mathcal{G} \setminus \{g\}} |\mathbf{h}_{b_u,k} \mathbf{w}_{u}|^{2}}}
\end{equation}
and the data rate towards user $k$ is given as
\begin{equation}
R_k(\mathbf{w}) \triangleq \log(1+\Gamma_k(\mathbf{w})).
\end{equation}

\subsection{Power Consumption Model}

The total power consumption at the transmitter side is \cite{Tervo-15}
\begin{equation}
\label{eq:powermodel}
\begin{aligned}
P_{\text{tot}}= & \dfrac{1}{\eta}\sum\limits_{g\in\mathcal{G}}||\mathbf{w}_{g}||_2^{2} +P_{\text{RF}}\sum_{b\in\mathcal{B}}\sum_{i \in \mathcal{N}_b}a_{b,i} + P_{\text{sta}}
\end{aligned}
\end{equation}
where $a_{b,i}\in\{0,1\}$ is the antenna selection variable for the $i$th transmit antenna of BS $b$, i.e.,
$a_{b,i}=1$ if the $i$th antenna is selected and $a_{b,i}=0$ otherwise, and $\eta\in[0,1]$ is the power amplifier
efficiency. Note that the first term in \eqref{eq:powermodel} is the power consumption of the PA's to get the desired output powers,
$P_{\text{RF}}$ is the power consumption of all circuit blocks in each active RF chain, $P_{\text{sta}}$
is the static power spent by cooling systems, power supplies, etc.

\subsection{Problem Formulation }
The optimization target is to maximize network energy efficiency
\begin{subequations} \label{EEmax}
\begin{eqnarray}{} \underset{\mathbf{w},\mathbf{a}}{\maxi} &  & \frac{\sum\limits_{g\in\mathcal{G}}\underset{k\in\mathcal{K}_g}{\min}\log(1+\Gamma_k(\mathbf{w}))}{\sum\limits_{g\in\mathcal{G}}\frac{1}{\eta}||\mathbf{w}_{g}||_2^{2} + P_{\text{RF}}\sum\limits_{b\in\mathcal{B}}\sum\limits_{i\in\mathcal{N}_b}a_{b,i} + P_{\text{sta}} }  \label{eq:EE:obj} \\ \st
 &   & ||\hat{\mathbf{w}}_{b,i}||_2^2 \leq a_{b,i}P_{\text{max}},  \forall b \in \mathcal{B}, i \in \mathcal{N}_b\label{eq:IntPconstraint}\\
 &   & \Gamma_k(\mathbf{w}) \geq \bar{\Gamma}_k,  \forall k \in \mathcal{K}, \label{eq:SINRconstraints}\\
 &   & a_{b,i} \in \{0,1\}, \forall b \in \mathcal{B}, i \in \mathcal{N}_b
\end{eqnarray}
\end{subequations}
where $\bar{\Gamma}_k$ is the SINR threshold for user $k$, $\mathbf{w}\triangleq\{\mathbf{w}_g\}_{g\in\mathcal{G}},\mathbf{a}\triangleq\{a_{b,i}\}_{b\in\mathcal{B},i\in\mathcal{N}_b}$, $\hat{\mathbf{w}}_{b,i} \triangleq [\mathbf{w}_{j}[i], \mathbf{w}_{j+1}[i],\ldots,\mathbf{w}_{j+G_b-1}[i]]$ includes the beamforming coefficients related to antenna $i$ of BS $b$ and index $j$ refers to a group index served by BS $b$. In the numerator of the objective \eqref{eq:EE:obj}, the rate for user group $g$ is defined as a minimum of the user rates in the group, because common information is transmitted to the group. The constraint in \eqref{eq:IntPconstraint} guarantees that if $a_{b,i}=0$, then the beamformers associated with antenna $i$ are set to zero. It also sets the maximum antenna specific power constraint.
The above problem is a non-convex mixed-Boolean fractional program which is hard to tackle as such. One of the main challenges is that the problem is non-convex even when the binary variables are relaxed to continuous. More specifically, in that case, \eqref{eq:SINRconstraints} and the numerator of the objective function are non-convex.

\section{Proposed Solution}\label{sec:CentralizedMethods}

\subsection{Equivalent Transformation}
To find a more tractable reformulation, we first equivalently transform \eqref{EEmax} as
\begin{subequations}
\label{eq:EEmax:reform1}
\begin{eqnarray}
\hspace{-10pt} \underset{\mathbf{w},\boldsymbol\gamma, \mathbf{v},\mathbf{a},\mathbf{r}}{\maxi} &  &  \frac{\sum_{g\in\mathcal{G}}r_g}{\sum\limits_{b\in\mathcal{B}}\sum\limits_{i\in\mathcal{N}_b}\frac{1}{\eta}v_{b,i} + P_{\text{RF}}\sum\limits_{b\in\mathcal{B}}\sum\limits_{i\in\mathcal{N}_b}a_{b,i} + P_{\text{sta}} }\\
\st
&  & \hspace{-10pt} ||\hat{\mathbf{w}}_{b,i}||_2^2 \leq a_{b,i}^\alpha v_{b,i},\; \forall b \in \mathcal{B}, i \in \mathcal{N}_b \label{eq:EEmax:reform1:MaxPC}\\
&  & \hspace{-10pt} v_{b,i} \leq P_{\text{max}}, \forall b \in \mathcal{B}, i \in \mathcal{N}_b \label{eq:EEmax:reform1:vmin}\\
&  & \hspace{-10pt}a_{b,i} \in \{0,1\}, \forall b \in \mathcal{B}, i \in \mathcal{N}_b\\
&  & \hspace{-10pt} \gamma_k \leq \frac{|\mathbf{h}_{b_g,k} \mathbf{w}_{g}|^{2}}{{\sigma_{k}^{2} + \sum\limits_{u \in \mathcal{G} \setminus \{g\}} |\mathbf{h}_{b_u,k} \mathbf{w}_{u}|^{2}}}, \forall k \in \mathcal{K}\label{eq:EEmax:reform1:rate}\\
&  & \hspace{-10pt} \gamma_k \geq \bar{\Gamma}_k, \forall k \in \mathcal{K}\label{eq:EEmax:reform1:minSINR}\\
&  & \hspace{-10pt} r_g \leq \log(1+\gamma_k), \forall g \in \mathcal{G}, k \in \mathcal{K}_g\label{eq:EEmax:reform1:weakestRATE}
\end{eqnarray}
\end{subequations}
where $\boldsymbol\gamma\triangleq\{\gamma_k\}_{k\in\mathcal{K}}, \mathbf{v}\triangleq\{v_{b,i}\}_{b\in\mathcal{B},i\in\mathcal{N}_b}, \mathbf{r}\triangleq \{r_g\}_{g\in\mathcal{G}}$ are new variables representing the SINR of each user $k$, the power transmitted from antenna $i$ of BS $b$ \cite{Tervo-15}, and the rate of the weakest link in user group $g$, respectively. It is not difficult to see that \eqref{EEmax} and \eqref{eq:EEmax:reform1} are equivalent, because the constraints in \eqref{eq:EEmax:reform1:rate} and \eqref{eq:EEmax:reform1:MaxPC} are active and $r_g = \underset{k\in\mathcal{K}_g}{\min}\log(1+\gamma_k), \forall g \in \mathcal{G}$ in \eqref{eq:EEmax:reform1:weakestRATE} is satisfied at the optimum. We have equivalently replaced $r_g\leq\underset{k\in\mathcal{K}_g}{\min}\log(1+\Gamma_k(\mathbf{w}))$ with $r_g\leq\log(1+\Gamma_k(\mathbf{w})), \forall k \in \mathcal{K}_g$ in \eqref{eq:EEmax:reform1:weakestRATE}, because if the minimum user rate satisfies the constraint, then all the user rates in that group have to satisfy it.  In constraint \eqref{eq:EEmax:reform1:MaxPC}, we have introduced a new formulation $||\hat{\mathbf{w}}_{b,i}||_2^2 \leq a_{b,i}^\alpha v_{b,i}$ which is equivalent for any value of $\alpha$ because each $a_{b,i}$ is binary.
  The exponent $\alpha\geq 1$ in \eqref{eq:EEmax:reform1:MaxPC} can be interpreted as a penalty parameter which penalizes the values of $a_{b,i}$ so that they are encouraged towards binary solution in case of continuous relaxation. The new formulation \eqref{eq:EEmax:reform1:MaxPC} reduces the feasible set of the problem when continuous relaxation is performed, so that the larger $\alpha$, the smaller feasible set.
In other words, we have the following inequalities
\begin{equation}
\text{EE}_{\text{bin}} \leq \text{EE}_{\text{cont,$\alpha>1$}} \leq \text{EE}_{\text{cont,$\alpha=1$}} \leq \text{EE}_{\text{cont,orig}}
\end{equation}
where $\text{EE}_{\text{bin}}, \text{EE}_{\text{cont,$\alpha>1$}}, \text{EE}_{\text{cont,$\alpha=1$}}$ and $\text{EE}_{\text{cont,orig}}$ are the optimal objective of Boolean formulation \eqref{EEmax}, continuous relaxation of \eqref{eq:EEmax:reform1} with $\alpha>1$, continuous relaxation of \eqref{eq:EEmax:reform1} with $\alpha=1$  and continuous relaxation of \eqref{EEmax}. In fact, the proposed formulation is a generalization of the approach used in \cite{Tervo-15} for single-cell multiuser MISO unicasting, where $\alpha=1$ and $\alpha=2$ were used. However, the numerical results show that the value of $\alpha$ has a significant effect on the algorithm performance, and, thus, the energy efficiency.

\subsection{Solving the Equivalent Transformation}
By taking a look at \eqref{eq:EEmax:reform1}, we observe that if the binary variables are relaxed as $0 \leq a_{b,i} \leq 1$, the objective function is a concave-convex fractional function and the difficulty in solving \eqref{eq:EEmax:reform1} is due to the constraints \eqref{eq:EEmax:reform1:rate} and \eqref{eq:EEmax:reform1:MaxPC}. By using the continuous relaxation and introducing new variables $\boldsymbol\beta \triangleq \{\beta_k\}_{k\in\mathcal{K}}$ to represent the total interference-plus-noise of user $k$, the problem becomes
\begin{subequations}
\label{eq:EEmax:reform2}
\begin{eqnarray}
\hspace{-15pt} \underset{\mathbf{w},\boldsymbol\gamma, \mathbf{v},\mathbf{a},\boldsymbol\beta,\mathbf{r}}{\maxi} &  &  \frac{\sum_{g\in\mathcal{G}}r_g}{\sum\limits_{b\in\mathcal{B}}\sum\limits_{i\in\mathcal{N}_b}\frac{1}{\eta}v_{b,i} + P_{\text{RF}}\sum\limits_{b\in\mathcal{B}}\sum\limits_{i\in\mathcal{N}_b}a_{b,i} + P_{\text{sta}} }\\
\st
&  &  \hspace{-15pt} ||\hat{\mathbf{w}}_{b,i}||_2^2 \leq a_{b,i}^\alpha v_{b,i},\; \forall b \in \mathcal{B}, i \in \mathcal{N}_b \label{eq:EEmax:reform2:MaxPC}\\
&  & \hspace{-15pt} 0 \leq a_{b,i} \leq 1, \forall b \in \mathcal{B}, i \in \mathcal{N}_b\\
&  & \hspace{-15pt} \gamma_k \leq \frac{|\mathbf{h}_{b_g,k} \mathbf{w}_{g}|^{2}}{\beta_k}, \forall k \in \mathcal{K} \label{eq:EEmax:reform2:quadoverlin}\\
&  & \hspace{-15pt} \beta_k \geq {\sigma_{k}^{2} + \sum\limits_{u \in \mathcal{G} \setminus \{g\}} |\mathbf{h}_{b_u,k}\herm \mathbf{w}_{u}|^{2}}, \forall k \in \mathcal{K}\label{eq:EEmax:reform2:betaconst}\\
&  & \hspace{-15pt} \eqref{eq:EEmax:reform1:MaxPC}, \eqref{eq:EEmax:reform1:vmin}, \eqref{eq:EEmax:reform1:minSINR}, \eqref{eq:EEmax:reform1:weakestRATE}.
\end{eqnarray}
\end{subequations}
Now the left side of \eqref{eq:EEmax:reform2:quadoverlin} is linear and the right side is convex quadratic-over-linear function.  To deal with the right side of \eqref{eq:EEmax:reform2:quadoverlin}, we can write its linear lower bound approximation at point $(\mathbf{w}_{g}^{(n)},\beta_{k}^{(n)})$ as
\begin{eqnarray}\label{eq:EEmax:quad_over_lin_app}
 |\mathbf{h}_{b_g,k}\mathbf{w}_{g}|^2/\beta_{k} \geq 2\Re((\mathbf{w}_{g}^{(n)})\herm\mathbf{h}_{b_g,k}\herm\mathbf{h}_{b_g,k}\mathbf{w}_{g})/\beta_{k}^{(n)}\nonumber \\
 - (|\mathbf{h}_{b_g,k}\mathbf{w}_{g}^{(n)}|/\beta_{k}^{(n)})^2\beta_{k} \triangleq \Psi_k^{(n)}(\mathbf{w}_{g},\beta_{k}).
\end{eqnarray}
To find a more tractable formulation for \eqref{eq:EEmax:reform2:MaxPC}, we first equivalently write it as
\begin{equation}
\frac{||\hat{\mathbf{w}}_{b,i}||_2^2}{v_{b,i}} \leq a_{b,i}^\alpha ,\; \forall b \in \mathcal{B}, i \in \mathcal{N}_b
\end{equation}
where the left side is a convex quadratic-over-linear function and the right side is convex. We
 can write the linear lower bound approximation of the right side at point $a_{b,i}^{(n)}$ as
\begin{eqnarray}\label{eq:EEmax:Approx}
 a_{b,i}^\alpha \geq (1-\alpha)(a_{b,i}^{(n)})^\alpha + \alpha {(a_{b,i}^{(n)})}^{(\alpha-1)}a_{b,i} \triangleq \chi_{b,i}^{(n)}(a_{b,i}).
\end{eqnarray}
With the approximations \eqref{eq:EEmax:quad_over_lin_app} and \eqref{eq:EEmax:Approx}, the problem becomes
\begin{subequations}
\label{eq:EEmax:reform3}
\begin{eqnarray}
\hspace{-20pt} \underset{\mathbf{w},\boldsymbol\gamma,\mathbf{v},\mathbf{a},\boldsymbol\beta,\mathbf{r}}{\maxi} &  &  \frac{\sum_{g\in\mathcal{G}}r_g}{\sum\limits_{b\in\mathcal{B}}\sum\limits_{i\in\mathcal{N}_b}\frac{1}{\eta}v_{b,i} + P_{\text{RF}}\sum\limits_{b\in\mathcal{B}}\sum\limits_{i\in\mathcal{N}_b}a_{b,i} + P_{\text{sta}} }\\
\st
&  & \frac{||\hat{\mathbf{w}}_{b,i}||_2^2}{v_{b,i}} \leq \chi_{b,i}^{(n)}(a_{b,i}), \forall b \in \mathcal{B}, i \in \mathcal{N}_b \label{eq:EEmax:reform3:ApproxB}\\
&  & \gamma_k \leq \Psi_k^{(n)}(\mathbf{w}_{g},\beta_{k}), \forall k \in \mathcal{K}\\
&  & 0 \leq a_{b,i} \leq 1, \forall b \in \mathcal{B}, i \in \mathcal{N}_b\\
&  & \eqref{eq:EEmax:reform1:MaxPC}, \eqref{eq:EEmax:reform1:vmin}, \eqref{eq:EEmax:reform1:minSINR},\eqref{eq:EEmax:reform2:betaconst}, \eqref{eq:EEmax:reform1:weakestRATE}.
\end{eqnarray}
\end{subequations}
At this point, we note that the problem is a concave-convex fractional program which can be transformed to a convex one with the Charnes-Cooper transformation \cite{Schaible-76}. Thus, solving \eqref{eq:EEmax:reform3} is equivalent to the following convex problem
\begin{subequations}
\label{eq:EEmax:reform4}
\begin{eqnarray}
& \hspace{-20pt} \underset{\phi,\bar{\mathbf{w}},\bar{\boldsymbol\gamma},\bar{\mathbf{v}},\bar{\mathbf{a}},\bar{\boldsymbol\beta},\bar{\mathbf{r}}}{\maxi} &   \sum_{g\in\mathcal{G}}\bar{r}_g\label{eq:EEmax:reform4:obj}\\
& \hspace{-60pt} \st  & \hspace{-35pt} \sum\limits_{b\in\mathcal{B}}\sum\limits_{i\in\mathcal{N}_b}\frac{1}{\eta}\bar{v}_{b,i} + P_{\text{RF}}\sum\limits_{b\in\mathcal{B}}\sum\limits_{i\in\mathcal{N}_b}\bar{a}_{b,i} + \phi P_{\text{sta}} \leq 1  \\
&  & \hspace{-35pt} \frac{||\bar{\hat{\mathbf{w}}}_{b,i}||_2^2}{\bar{v}_{b,i}} \leq \phi\chi_{b,i}^{(n)}(\frac{\bar{a}_{b,i}}{\phi}), \forall b \in \mathcal{B}, i \in \mathcal{N}_b\\
&  & \hspace{-35pt} \bar{v}_{b,i} \leq \phi P_{\text{max}}, \forall b \in \mathcal{B}, i \in \mathcal{N}_b \\
&  & \hspace{-35pt} 0 \leq \bar{a}_{b,i} \leq \phi, \forall b \in \mathcal{B}, i \in \mathcal{N}_b\\
&  & \hspace{-35pt} \bar{\gamma}_k \leq \phi\Psi_k^{(n)}(\frac{\bar{\mathbf{w}}_{g}}{\phi},\frac{\bar{\beta}_{k}}{\phi}), \forall k \in \mathcal{K}\\
&  & \hspace{-35pt} \bar{\gamma}_k \geq \phi \bar{\Gamma}_k, \forall k \in \mathcal{K} \\
&  & \hspace{-35pt} \phi\bar{\beta}_k \geq {\phi^2 \sigma_{k}^{2} + \sum\limits_{u \in \mathcal{G} \setminus \{g\}} |\mathbf{h}_{b_u,k}\herm \bar{\mathbf{w}}_{u}|^{2}}, \forall k \in \mathcal{K}\\
&  & \hspace{-35pt} \bar{r}_g \leq \phi\log(1+\frac{\bar{\gamma}_k}{\phi}), \forall g \in \mathcal{G}, k \in \mathcal{K}_g.
\end{eqnarray}
\end{subequations}
After solving \eqref{eq:EEmax:reform4}, the optimal solutions for the original fractional program \eqref{eq:EEmax:reform3} can be found as $\mathbf{w}_g^* = \bar{\mathbf{w}}_g^*/\phi^*, \beta_k^* = \bar{\beta}_k^*/\phi^*,\gamma_k^* = \bar{\gamma}_k^*/\phi^*,a_{b,i}^* = \bar{a}_{b,i}^*/\phi^*,v_{b,i}^* = \bar{v}_{b,i}^*/\phi^*,r_{g}^* = \bar{r}_{g}^*/\phi^*$, where $\bar{\mathbf{w}}_g^*,\phi^*, \bar{\beta}_k^*,\bar{\gamma}_k^*,\bar{a}_{b,i}^*,\bar{v}_{b,i}^*,\bar{r}_{g}^*$ are the optimal variables of \eqref{eq:EEmax:reform4}.
As a result, we use the idea of successive convex approximation \cite{Beck:SCA:2010}, where the convex problem \eqref{eq:EEmax:reform4} is solved at iteration $n$. After solving the problem at iteration $n$, the optimal solutions $\mathbf{w}_g^*, \beta_k^*, a_{b,i}$ are then used to update $\Psi_k^{(n+1)}(\mathbf{w}_{g},\beta_{k})$ and $\chi_{b,i}^{(n+1)}(a_{b,i})$ for the next iteration. The monotonic convergence of the objective function \eqref{eq:EEmax:reform4:obj} is not difficult to see, and a more detailed convergence analysis for the problem with similar structure can be found, e.g., in \cite[Appendix A]{Venkatraman-16}.

\begin{rem}
If at least one of the SINR targets $\bar{\Gamma}_k$ in some user group of BS $b$ is non-zero, we can further reduce the feasible set of \eqref{eq:EEmax:reform4} by adding the constraints
\begin{equation}
\sum_{i\in\mathcal{N}_b}a_{b,i} \geq \phi X_b, \forall b \in \mathcal{B}
\end{equation}
where $X_b$ is the number of groups served by BS $b$ which have at least one user having non-zero SINR target. This can be done because it is known that at least $X_b$ antennas have to be active to be able to transmit $X_b$ independent data streams.
\end{rem}

\subsubsection*{Recovering the Binary Solution from Continuous Relaxation}
Generally, solving the continuous relaxation usually results in a solution where many of the antenna selection variables are non-binary.
 However, due to the new formulation in \eqref{eq:EEmax:reform2:MaxPC}, many of the continuous antenna selection variables converge to binary. Thus, we propose to switch off all the antennas for which $a_{b,i}<\epsilon$, where $\epsilon$ is a small threshold. After performing the antenna selection, the algorithm needs to be run again with the selected antenna set to find the beamformers with lower dimensions. However, in the numerical results, we also illustrate that the beamformers returned by the relaxed problem already yields a good energy efficiency with the correct choice of $\alpha$. This method is called `Alg. 1 \emph{`simple'}' in the numerical results. The proposed joint beamforming and antenna selection method is summarized in Algorithm \ref{algo:iterative}.
\subsubsection*{Initial Points}
One of the challenges is to find feasible initial points especially due to the QoS constraints. The initial $\mathbf{a}^{(0)}$ can be set to all-ones.
To find $\mathbf{w}^{(0)},\boldsymbol\beta^{(0)}$, we can apply a similar approach as that in \cite{Vu-16}. We first initialize any $\mathbf{w}^{(0)},\boldsymbol\beta^{(0)}$ and then consider a penalized formulation of \eqref{eq:EEmax:reform4}, which can be written in a general form as
\begin{subequations}
\label{eq:EEmax:feas}
\begin{eqnarray}
  \underset{\mathbf{x}}{\max}  &  &  f_0(\mathbf{x})-\lambda||\mathbf{q}|| \\
  \st & &
   f_i(\mathbf{x}) \leq q_i, i=1,\ldots,Q_1 \\
&  &   F_j(\mathbf{x},\mathbf{x}^{(n)}) \leq q_j, j=1,\ldots,Q_2
\end{eqnarray}
\end{subequations}
where $f_0(\mathbf{x})$ is a concave objective function, $f_i(\mathbf{x})$ and $F_j(\mathbf{x},\mathbf{x}^{(n)})$ are convex constraints, $\mathbf{x}$ includes all the variables in \eqref{eq:EEmax:reform4}, $\lambda$ is a positive penalty parameter and $\mathbf{q} \in \mathcal{R}_+^{Q_1+Q_2}$ are new slack variables, one for each constraint in \eqref{eq:EEmax:reform4}. The above problem is iteratively solved by updating the constraints $F_j(\mathbf{x},\mathbf{x}^{(n)})$ (i.e., $\Psi_k^{(n)}(\frac{\bar{\mathbf{w}}_{g}}{\phi},\frac{\bar{\beta}_{k}}{\phi})$ in \eqref{eq:EEmax:reform4}) after each iteration, until $q_i = 0, \forall i$. The feasible point is found very efficiently because the elements of $\mathbf{q}$ are encouraged to zero due to the penalty function in the objective.

\begin{algorithm}[t]
\begin{algorithmic}[1] \caption{Proposed joint beamforming and antenna selection design.}
\label{algo:iterative} \renewcommand{\algorithmicrequire}{\textbf{Initialization:}}
\REQUIRE Set $n=0$, and generate feasible initial points $(\mathbf{w}^{(n)},\boldsymbol\beta^{(n)},\mathbf{a}^{(n)})$.
\REPEAT
\STATE Solve \eqref{eq:EEmax:reform4} with $(\mathbf{w}^{(n)},\boldsymbol\beta^{(n)},\mathbf{a}^{(n)})$
and denote optimal values as $(\bar{\mathbf{w}}^*,\bar{\boldsymbol\beta}^*,\bar{\mathbf{a}}^*)$.
\STATE Update $\mathbf{w}^{(n+1)} = \bar{\mathbf{w}}^*/\phi ,\boldsymbol\beta^{(n+1)} = \bar{\boldsymbol\beta}^*/\phi,\mathbf{a}^{(n+1)}=\bar{\mathbf{a}}^*/\phi$ and $\chi_{b,i}^{(n+1)}(a_{b,i}), \Psi_k^{(n+1)}(\mathbf{w}_{g},\beta_{k})$.\label{algo:update}
\STATE $n:=n+1$.
\UNTIL convergence
\renewcommand{\algorithmicrequire}{\textbf{Output:}} \REQUIRE $a_{b,i}^* = \tfrac{\bar{a}_{b,i}^*}{\phi^*}, \forall b \in \mathcal{B}, i \in \mathcal{N}_b$
\STATE Set $a_{b,i}=0$, for all $b,i$ for which $a_{b,i}^*<\epsilon$.
\STATE Run steps 1-5 again with fixed $\mathbf{a}$ to find beamformers with reduced dimensions.
\renewcommand{\algorithmicrequire}{\textbf{Output:}} \REQUIRE $\mathbf{w}_{g}^* = \tfrac{\bar{\mathbf{w}}_{g}^*}{\phi^*}, \forall g \in \mathcal{G}$
\end{algorithmic}
\end{algorithm} 

\section{Numerical Results}
\label{sec:NumericalResults}

We evaluate the performance for a quasistatic frequency flat Rayleigh fading channel model with $B=2$ base stations. Each BS serves two groups of users with $|\mathcal{K}_g|=L=2$ users per group, i.e., the total number of users in the network is 8. The worst-case interference scenario is assumed so that the average path loss from all the BSs to all the users is 0 dB. We assume a unit bandwidth and noise power is normalized to $\sigma_k^2=0$ dBW, $\forall k\in\mathcal{K}$. We set $N_b = N$ for all $b$, i.e., $N$ is the number of antennas at each BS, and the algorithms have been stopped when the change of the objective value has been smaller than $\xi=10^{-6}$ between two last iterations. We set $\eta=0.35, \epsilon=0.001, P_{\text{max}}=9$ dBW, $P_{\text{sta}}=2$ W, $\bar{\Gamma}_k=\bar{\Gamma}, \forall k \in \mathcal{K}$, and the other simulation parameters are given in the figures.

\begin{figure}[t]
\centering
  \includegraphics[width=0.79\columnwidth]{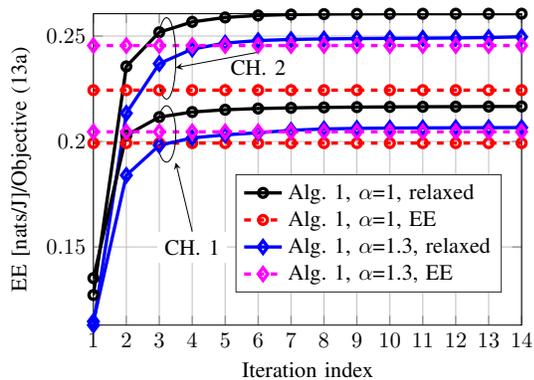}
  \caption{Convergence of the relaxed problem and achieved energy efficiency for two different channel realizations with $P_{\text{RF}}=$1 W, $N=$16, $\bar{\Gamma}=$ 0 dB. The flat lines denote the achieved energy efficiencies after calculating the beamformers for the chosen antenna sets (i.e., after terminating Alg. \ref{algo:iterative}).}
  \label{fig: Convergence}
\end{figure}

Fig. \ref{fig: Convergence} illustrates the convergence of the relaxed problem and the achieved energy efficiency for two different values of $\alpha$, and two random channel realizations (denoted as `CH' in the figure). The same initial points have been used for both values of $\alpha$. We can see that the convergence speed is fast in the considered settings for both cases. However, we observe that they converge to different solutions in both examples. Specifically, the objective value of the relaxed problem after convergence is higher for $\alpha=1$, but the achieved energy efficiency (after recovering the binary solution) is worse than with $\alpha=1.3$. This is because with $\alpha=1$, more antenna selection variables are non-binary after convergence, which results in worse antenna selection result. Another observation is that with $\alpha=1.3$ (which achieves better energy efficiency), the objective value returned by the relaxed problem and the achieved energy efficiency are very close to each other. This means that the solution of the relaxed problem is already very close to binary. The examples demonstrate the effectiveness of the proposed formulation. The impact of $\alpha$ on the average energy efficiency is studied in the next experiment.

\begin{figure}[t]
\centering
  \includegraphics[width=0.79\columnwidth]{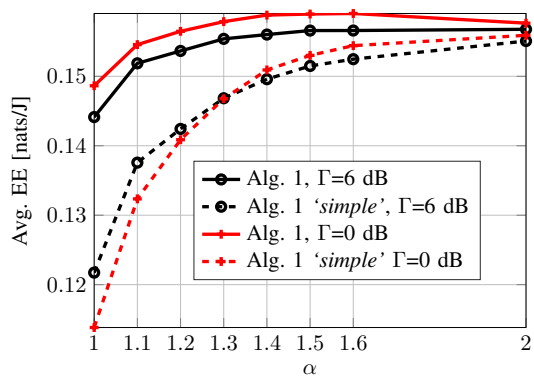}
  \caption{Energy efficiency versus $\alpha$ with $P_{\text{RF}}=$2 W, $N=$16.}
  \label{fig: EEvsAlpha}
\end{figure}

\begin{figure}[t]
\centering
  \includegraphics[width=0.79\columnwidth]{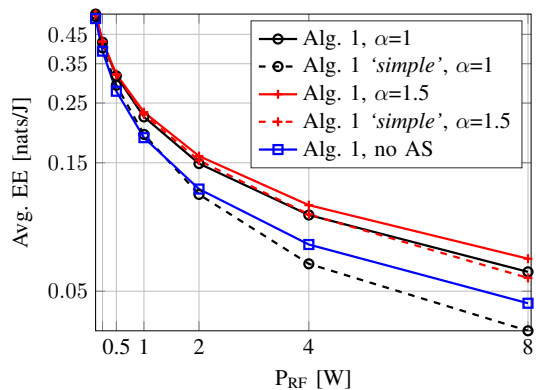}
  \caption{Energy efficiency versus $P_{\text{RF}}$ with $N=$16, $\bar{\Gamma}=$ 0 dB.}
  \label{fig: EEvsPdyn}
\end{figure}

\begin{figure}[t]
\centering
  \includegraphics[width=0.79\columnwidth]{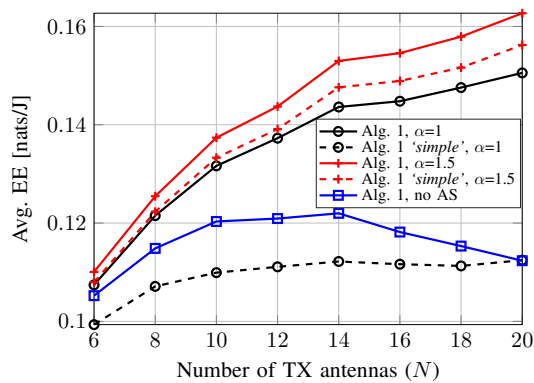}
  \caption{Energy efficiency versus $N$ with $P_{\text{RF}}=$2 W, $\bar{\Gamma}=$ 0 dB.}
  \label{fig: EEvsN}
\end{figure}

Fig. \ref{fig: EEvsAlpha} illustrates the effect of $\alpha$ on the average energy efficiency for $\bar{\Gamma}=$ 0 dB and $\bar{\Gamma}=$ 6 dB. We also illustrate the performance of the simplified algorithm (Alg. 1 \emph{'simple'}), where the beamformers achieved from the relaxed problem (i.e., step 7 is ignored in Alg. 1) are used for transmission. We can see that the choice of $\alpha$ affects the achieved energy efficiency, and the choice of the best $\alpha$ depends on the system parameters. In this case, $\alpha=1.6$ gives the best performance when $\bar{\Gamma}=$ 0 dB, and for larger $\bar{\Gamma}$, it is better to use larger $\alpha$. The reason for this is that with larger $\bar{\Gamma}$, the transmit (TX) power has to be higher to satisfy the SINR constraints. As a result, in the relaxed problem, also the relaxed variables $a_{b,i}$ have to be larger, implying that the larger 'penalty' (i.e. $\alpha$) is required to force them to zero. We can also see that the larger the $\alpha$, the better results are achieved with the simple method, implying that the relaxed problem yields near-binary solutions. This again demonstrates the benefit of the proposed formulation.

Fig. \ref{fig: EEvsPdyn} illustrates the average energy efficiency versus $P_{\text{RF}}$. Specifically, the proposed algorithm is run with $\alpha=1$ and $\alpha=1.5$, and compared with the method where only beamformers are optimized (Alg. 1, no AS). First, we can observe that Alg. 1 with $\alpha=1.5$ achieves the best results, and the gap between $\alpha=1$ and  $\alpha=1.5$ increases with $P_{\text{RF}}$. It is interesting to observe that with $\alpha=1.5$, even the beamformers obtained from the relaxed problem result in very good performance, which again verifies that the relaxed problem yields near-binary solutions. However, the simple method with $\alpha=1$ is even worse than the method without antenna selection for larger $P_{\text{RF}}$. Finally, we can see that the proposed algorithm provides significant energy efficiency gains over the method without antenna selection (i.e., roughly 5-50\% with the considered system parameters).

Fig. \ref{fig: EEvsN} plots the average EE versus the number of antennas per BS. First, we can see that EE starts to decrease without antenna selection when $N>14$. As can be observed, significant gains are achieved with the proposed algorithm and the gains increase with the number of antennas. Once again, significant gains are achieved with $\alpha=1.5$ compared to the method with $\alpha=1$. In addition, the simple method with $\alpha=1.5$ gives even better performance than the original algorithm with $\alpha=1$ and the gain increases with $N$. The simple method with $\alpha=1$, on the other hand, yields poor performance.

\begin{figure}[t]
\centering
  \includegraphics[width=0.79\columnwidth]{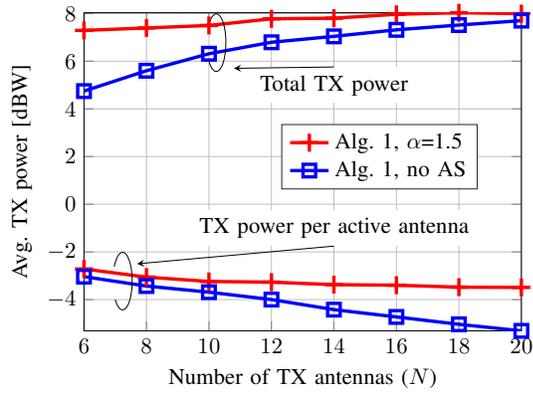}
  \caption{Total transmit power and per-antenna power versus $N$ with $P_{\text{RF}}=$2 W, $\bar{\Gamma}=$ 0 dB.}
  \label{fig: TXpowervsN}
\end{figure}

Fig. \ref{fig: TXpowervsN} illustrates the transmit powers versus $N$ with the proposed algorithm and a method without antenna selection. We can see that it is energy-efficient to increase sum power when the number of antennas increases. However, the average transmit power per active antenna decreases with increasing $N$ also with antenna selection. This means that the more available antennas, the more antennas are chosen to maximize the energy efficiency.

\section{Conclusions}
\label{sec:Conclusions}
This paper has studied energy-efficient multi-cell multigroup coordinated joint beamforming and antenna selection with antenna-specific maximum power constraints and user-specific QoS constraints. The resulting mixed-Boolean nonconvex optimization problem was tackled by continuous relaxation and successive convex approximation, where the antennas for which continuous antenna selection variables converge to zero are switched off. The numerical results have illustrated the achieved energy efficiency gains of the proposed methods over the method without antenna selection.

\section{Acknowledgements}
This work was supported in part by Infotech Oulu Doctoral Program and the Academy of Finland under projects MESIC belonging to the WiFIUS program with NSF, and WiConIE. It was also supported by a research grant from Science Foundation Ireland and is co-funded by the European Regional Development Fund under Grant 13/RC/2077, projects FNR SEMIGOD, SATSENT, INWIPNET and H2020 SANSA. The first author has been supported by KAUTE Foundation, HPY Research Foundation, Walter Ahlström Foundation, Tauno Tönning Foundation, and Nokia Foundation.


\end{document}